\documentclass[12pt,a4paper]{article}

\usepackage{amssymb}
\usepackage[numbers]{natbib}

\usepackage[figuresright]{rotating}
\usepackage{algorithm} 
\usepackage{amsmath}
\usepackage{algpseudocode} 
\usepackage{graphicx}
\usepackage{xcolor}
\usepackage[utf8]{inputenc}
\usepackage[greek,english]{babel}
\usepackage{lmodern}
\usepackage{pdflscape}
\usepackage{adjustbox}
\usepackage{tabularx}
\usepackage{array}
\usepackage{graphicx}
\usepackage{adjustbox}
\usepackage{float}
\usepackage{blindtext}
\usepackage{enumitem}
\usepackage{gensymb}
\usepackage{xcolor}
\usepackage{soulutf8}
\usepackage{caption} 
\usepackage{xcolor}
\usepackage{amsmath}
\usepackage{algorithm} 
\usepackage{algpseudocode} 
\usepackage{multicol}
\usepackage{wrapfig}

\begin{document}

\title{Evaluating and Combatting the Impact of Concept Drift on the Performance of Machine Learning-Based Phishing Detection Systems}

\author{Warren Fernando, Nikos Komninos \\
\small Department of Computer Science, School of Mathematics, Computer Science and Engineering, \\
\small City St George's, University of London, UK}

\date{}

\maketitle

\begin{abstract}
The expansion of the digital realm has caused a significant increase in digital communication, with emails being one of the most prominent forms of digital communication. The growth of email communication is applicable to professional and personal spaces, leaving many surface areas for attackers to exploit. Spam emails, a form of unsolicited emails that are often malicious to recipients, have been an ever-present problem for email users since the email's inception and the digital realm's expansion on amplifying this problem. Email spam filters are a staple in email clients, and they are designed to identify these potentially malicious emails and alert the recipient of their malicious nature. Phishing is often the first step in malware-based attacks, and the fast-paced evolution of malware means that phishing attacks also get more sophisticated over time. A common solution to detecting malicious behaviour in the malware and spam domain is machine learning. We attempt to measure how evolution in the spam email space affects these machine-learning-based detection systems and how degradation in performance can be reduced.
\end{abstract}


\section{Introduction}

Phishing is a type of social engineering that attempts to steal personal or sensitive information from victims by using deceptive emails or messages [1]. Affected victims may be tricked into clicking on URLs that can trick them into handing over sensitive information or opening files, which may compromise their personal information. The importance of phishing detection continues to rise because a large portion of personal and business transactions happen online; this gives attackers a large scope for deceiving victims [2]. Phishing is a constantly evolving and developing attack vector, with the number of phishing websites increasing from around 6.9 million in 2020 to around 14 million in 2023 [3]. Phishing originated in the 1990s, during the inception of dial-up Internet, and it gradually expanded into the 2000s. The first major phishing attack was the LoveBug phishing emails, which spread a worm that significantly damaged the host machine [4]. Phishing techniques have evolved over time to counteract the advancements in phishing and spam filters. Machine learning-based phishing detection methods have been widely used in phishing detection research and have yielded strong results; however, as phishing methods change and evolve, these models slowly become obsolete [5]; this phenomenon is known as concept drift in machine learning. The FeSAD framework has proven effective in the ransomware domain and has reduced the effect of concept drift on ransomware detection systems. We hypothesise that a similarity index combined with a classification algorithm can help improve classification performance under concept drift.
\subsection{Motivation}
The phishing attack vector is constantly evolving, and mathematical frameworks have been shown to be effective at combating concept drift in malware detection systems. The approach has been proven effective in the ransomware domain, and testing it on phishing data is a logical step, as this is also a constantly evolving attack vector. We hypothesise that we can use a statistical method to evaluate how similar or dissimilar phishing samples are to benign samples in prominent phishing datasets. Large phishing datasets enable us to evaluate which features are most effective for phishing classification and which provide the most robust platform for detecting phishing samples despite ongoing evolution.

\section{Introduction}
Phishing is a type of social engineering that attempts to steal personal or sensitive information from victims by using deceptive emails or messages [1]. Affected victims may be tricked into clicking on URLs that can trick them into handing over sensitive information or opening files, which may compromise their personal information. The importance of phishing detection continues to rise because a large portion of personal and business transactions happen online; this gives attackers a large scope for deceiving victims [2]. Phishing is a constantly evolving and developing attack vector, with the number of phishing websites increasing from around 6.9 million in 2020 to around 14 million in 2023 [3]. Phishing originated in the 1990s, during the inception of dial-up Internet, and it gradually expanded into the 2000s. The first major phishing attack was the LoveBug phishing emails, which spread a worm that significantly damaged the host machine [4]. Phishing techniques have evolved over time to counteract the advancements in phishing and spam filters. Machine learning-based phishing detection methods have been widely used in phishing detection research and have yielded strong results; however, as phishing methods change and evolve, these models slowly become obsolete [5]; this phenomenon is known as concept drift in machine learning. The FeSAD framework has proven effective in the ransomware domain and has reduced the effect of concept drift on ransomware detection systems. We hypothesise that a similarity index combined with a classification algorithm can help improve classification performance under concept drift.
\subsection{Motivation}
The phishing attack vector is constantly evolving, and mathematical frameworks have been shown to be effective at combating concept drift in malware detection systems. The approach has been proven effective in the ransomware domain, and testing it on phishing data is a logical step, as this is also a constantly evolving attack vector. We hypothesise that we can use a statistical method to evaluate how similar or dissimilar phishing samples are to benign samples in prominent phishing datasets. Large phishing datasets enable us to evaluate which features are most effective for phishing classification and which provide the most robust platform for detecting phishing samples despite ongoing evolution.

\subsection{Paper Contribution}
This paper's main contribution is the application of a statistical framework to a phishing detection scenario where we investigate phishing evolution; we theorise that the statistical approach can detect phishing attempts under concept drift and effectively extend the lifespan of a machine learning detection system.
\begin{itemize}
	
	\item \textbf{Statistical Approach for Use on Phishing Data: } We have utilised a statistical framework to analyse various phishing datasets and have induced concept drift scenarios to evaluate how well the framework can mitigate the effects of concept drift on M-L solutions designed to detect phishing.

	\item \textbf{Phishing Dataset Evaluation: } This framework derives a numeric value to quantify concept drift and makes reliable classifications of samples showing concept drift. We use these numeric values to evaluate the datasets we have used to determine what features provide the most distinguishable difference between benign URLs and phishing URLs used in phishing attempts. The drift values we use can determine the quality of the data and features by measuring the distinguishable difference between the average phishing and benign samples. Additionally, we assess what features remain the most robust to concept drift and remain important while phishing techniques may evolve.	
	
\end{itemize}

\subsection{Paper Organisation}

The remainder of this paper is as follows: Section 2 covers relevant work to this paper. Section 3 covers our framework and what we propose. Section 4 describes our experimental setup, and section 5 discusses the results of these experiments. Section 6 concludes our work and covers possible future work.

\section{Related Work \& Background}
\subsection{Phishing Detection with Machine Learning}
Machine learning methods have been widely used for phishing detection research, with some strong results being achieved. Alam et al. use random forests and decision trees to build classifiers to detect phishing URLs in [7]; this research achieves strong detection results of around 97\%. The work presented by Kosmopoulos et al. in [8] shows early evidence that Bayesian Networks can be used as an adaptive method for phishing and spam filters. The research presented in [9] by Saraswat et al. uses various machine learning algorithms to achieve promising results close to 100\% accuracy; however, the datasets used to test these algorithms are relatively small, with each dataset containing less than 600 samples. The work proposed by Saha et al. uses Deep Learning algorithms for phishing detection. Specifically, Saha et al. use a Multilayer Perceptron to build a deep-learning model to detect Phishing emails; however, similarly to Saraswat et al., the datasets used to validate and test these models are relatively small, with under 2000 samples being used. The main focus of phishing detection research is on building detection models; however, evolution in phishing is not at the forefront of building phishing detection models. Evolution in malware detection tasks presents as concept drift in machine-learning algorithms; the research proposed in [11] by Jordaney et al. addresses the problem of concept drift in machine-learning malware detection and proposes a detection and retraining approach that allows operators to inspect samples that show adverse evolutionary properties manually; samples are determined to show drift based on p-values and the prediction confidence of the underlying algorithm. The research proposed in [12] uses a hybrid machine-learning approach to detect phishing URLs and achieves promising results using the Microsoft Kaggle dataset. Lakshmanarao et al., explore using a hybrid methodology on the UCI dataset that combines multiple algorithms and uses an ensemble-type approach. The hybrid approach proposed in [13] achieves interesting results, with a detection rate of 97\%. The research proposed in [14] leverages various machine learning algorithms and determines that the Google Index of a URL plays a significant part in determining whether it is phishing or not; this is done using the Kaggle dataset. The research shown by Bouijij et al., tests various machine learning algorithms on phishing URL data and finds the K-Nearest Neighbour and Extra-Tree algorithm to be the most accurate of those tested. Chandra et al., achieve a detection rate of 98\% using the random forest algorithm on a dataset containing over 11,000 samples with an 80/20 split [15]. Bagui et al. use a vectorised dataset of email messages and test various machine-learning approaches' ability to vectorise phishing emails. The Word Embedding approach achieves an accuracy of 98.89\%, with a 70/30 training and test split with a dataset of 3416 phishing emails and 148950 benign emails. Chinnasamy et al., explore using machine learning techniques to detect phishing by focusing on the URL of the website the victim is directed to. The dataset used in [17] contains 1871 phishing and benign samples with a 70/30 training test split. The algorithms used achieve relatively modest results with detection rates between 70 and 30\%. Uplenchwar et al. also explore detecting phishing from text messages using multiple machine-learning algorithms; this research achieves detection results of 97\% and uses a dataset of 5572 text message samples [19]. Overall, multiple studies leverage machine learning algorithms to detect phishing attempts. These studies use different data types for training their models, with the most common approach being using the URLs to distinguish between legitimate and benign messages, whereas other approaches involve vectorising whole email messages to use as training data for phishing detection systems. These studies achieve strong results in both cases but do not consider evolution. Phishing methods have been known to change over time, with phishing scams becoming more and more advanced in the last two decades; additionally, with real-world phishing experiences, it is not uncommon for phishing emails to slip through modern spam filters as they are not necessarily set up to deal with evolution [20]. 

\subsection{Concept Drift}

Concept drift is best described as changes in $P(x, y)$. A prediction comprises of the feature probability component $P(x)$ and class label conditional probability $P(y|x)$. The changes in prediction probability can be understood through the changes in either of these two components. Concept Drift can be broken down into four types [6].

\begin{description}
	
	\item[$\cdot$ No Change]
	In this scenario, both $P(x)$ and $P(y|x)$ remain the same and have not changed regarding how they were when the model was initially trained. It might be useful to retrain on the most up-to-date data regardless; however, this indicates no concept drift, and the model is stable [6].
	
	\item[$\cdot$ Feature Change]
	In this scenario, $P(x)$ has changed, but $P(y|x)$ has remained the same; this means features that may not have been important have now become important. The model can be reconstructed in a scenario to fit these feature changes [6].
	
	\item[$\cdot$ Conditional Change]
	
	In this scenario, $P(x)$ has not changed, but $P(y|x)$ has changed. In this scenario, an expected error could go either way, so it is necessary to reconstruct the model. The issue with this is that there may not be enough new data to reconstruct the model, leading to high variance. It is possible to solve this problem through the weighted combination of new and old data [6].
	
	\item[$\cdot$ Dual Change]
	Both $P(x)$ and $P(y|x)$ change. The expected error could increase, decrease or stay the same depending on the combination of $P(x)$ and $P(y|x)$. It is necessary to retrain in this instance [6].
\end{description}

\subsection{Phishing Detection with Machine Learning with Concept Drift}
Evolution in malware detection tasks presents as concept drift in machine-learning algorithms; the research proposed in [21] by Jordaney et al. addresses the problem of concept drift in machine-learning malware detection and proposes a detection and retraining approach that allows operators to inspect samples that show adverse evolutionary properties manually; samples are determined to show drift based on p-values and the prediction confidence of the underlying algorithm. The research proposed in [22] uses a hybrid machine-learning approach to detect phishing URLs and achieves promising results using the Microsoft Kaggle dataset. In terms of concept drift research for phishing, there are some prominent research studies which attempt to address concept drift. The research explored in [23] uses Autoencoders to increase the detection rate under concept drift, and some positive results are achieved; however, the performance of the autoencoders appears inconsistent from test to test. The research shown in [24] addresses concept drift by calculating the difference between distributions and updating classifiers when drift is detected; however, the approach relies on updating and retraining classifiers instead of dealing with singular drifting examples.

\section{Our Proposal}

\begin{figure*}[t]

	\hspace{-3.4cm}\includegraphics[width=1.5\textwidth]{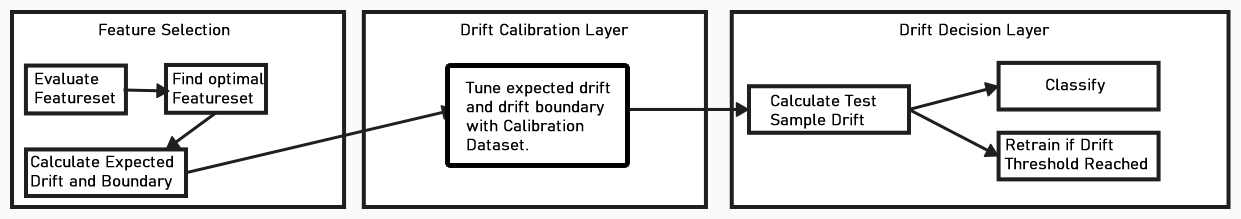}
	\captionof{figure}{Detection framework}
	\label{fig_1}
\end{figure*}
Figure 1 showcases the design of the Detection framework, which serves as the basis for this research. Our proposed approach uses a mathematical approach similar to the FeSAD framework [6], which is designed to be machine-learning algorithm-agnostic and aims to reduce the effects of concept drift on machine learning algorithms. The framework uses a user-defined machine-learning algorithm. The framework uses a combination of mathematical distance and similarity metrics, along with algorithmic prediction probabilities, to calculate sample "drift". Sample drift measures how similar or dissimilar a sample is compared to the other samples in the class in which it has been classified. The framework's approach aims to reduce the effects of malware evolution, which degrades machine learning-based malware classifiers. We hypothesise that a mathematical framework can be applied to phishing data and improve detection under concept drift. The mathematical approach we use has been shown to reduce the impact of ransomware evolution on machine learning-based ransomware classifiers. The mathematical framework takes the following approach: 
\subsection{Feature Selection}
The feature selection mechanism generates feature sets using a genetic algorithm [25]. The modified genetic algorithm that identifies high-information-gain features and incorporates them into all generated feature sets; this approach shows higher resistance to concept drift [25] and less degradation when the classifier is exposed to it. The feature selection process uses the training data to calculate drift values; this approach enables the framework to identify which drift levels correspond to misclassifications and which to correct classifications. Additionally, the feature selection mechanism calculates the average drift for correct classifications, enabling the framework to identify abnormal samples. The two key drift levels calculated are unacceptable drift and abnormal drift. Unacceptable drift would imply an incorrect prediction, and abnormal drift would imply a level of drift that was beyond the average, but it does not imply a misclassification.

\subsubsection{Distance Ratio}
The Distance Ratio represents the ratio of a sample's distance from malign samples to a samples distance to benign samples. The distance is measured using the respective similarity measure chosen by the user. The distance ratio metric measures the ratio of the distance of a sample to malign samples to the distance of a sample to benign samples. Eq.1 and 2 measure the average distance of a sample from phishing and benign samples, respectively, and Eq.3 is the ratio value of these two values. Benign samples will have a larger $dist_i$ than malign samples because their average distance to malign $dist_m$ should be significantly larger than their distance to benign samples $dist_b$.
\begin{align}
	dist_m = \frac{\sum_{i, j=1}^{n} d_{(x_{i_m},x_{j_m})}}{n}
\end{align}
\begin{align}
	dist_b = \frac{\sum_{i, j=1}^{n} d_{(x_{i_b},x_{j_m})}}{n}
\end{align}

\begin{align}
	dist_{i} = \frac{dist_m}{dist_b}
\end{align}

The Distance Ratio represents the ratio of a sample's distance from malign samples to a samples distance to benign samples. The distance is measured using the respective similarity measure chosen by the user. The distance ratio metric measures the ratio of the distance of a sample to malign samples to the distance of a sample to benign samples. Eq.1 and 2 measure the average distance of a sample from phishing and benign samples, respectively, and Eq.3 is the ratio value of these two values. Benign samples will have a larger $dist_i$ than malign samples because their average distance to malign $dist_m$ should be significantly larger than their distance to benign samples $dist_b$.
\begin{align}
	dist_m = \frac{\sum_{i, j=1}^{n} d_{(x_{i_m},x_{j_m})}}{n}
\end{align}
\begin{align}
	dist_b = \frac{\sum_{i, j=1}^{n} d_{(x_{i_b},x_{j_m})}}{n}
\end{align}

\begin{align}
	dist_{i} = \frac{dist_m}{dist_b}
\end{align}

\subsubsection{Prediction Weights}
The prediction weights are shown below in Eq.4 and 5. The weights are calculated by dividing the prediction probability of the sample by the average incorrect prediction for malign samples $\bar{p_{x_m}}$ for phishing and $\bar{p_{x_b}}$ for benign predictions. The average prediction probabilities for incorrect classifications are derived from the training set.

\begin{align}
	w_m= \frac{p_i}{\bar{p_{x_m}}}
\end{align}
\begin{align}
	w_b= \frac{p_i}{\bar{p_{x_b}}}
\end{align}

\subsection{Weighted Distance Calculation}
The weighted distance combines prediction probability and the metrics extracted from the distance or dissimilarity measure. The weighted distance calculation for samples classified as malign is shown in Eq.9. and the weighted distance for a sample classified as benign is shown in Eq.10. The weights are designed to shrink malign distance values if the prediction probability is high and increase benign distance values if the prediction probability is high, creating greater separation between samples when the algorithm is confident in its predictions and samples are statistically similar to its respective predicted class.
\begin{align}
	dw_{b_i}= \log (dist_{i} \cdot w_b)
\end{align}
\begin{align}
	dw_{m_i}= \log (\frac{dist_{i}}{w_m})
\end{align}

\subsubsection{Expected Drift Thresholds}
Our framework uses drift calculations that formally measure how far from an expected point a value is; The average weighted distance for correct malign classifications is $\bar{dw_{m_c}}$ and the average weighted distance for correct benign predictions is $\bar{dw_{b_c}}$. The drift calculations measure the difference between the average weighted distance in the training set and the weighted distance of an individual sample. The weighted distance for a sample $i$ is represented with $dw_{m_i}$ for malign and $dw_{b_i}$ for benign samples. The calculation for drift in a sample classified as malign is shown in equation 11, and the calculation for drift in a sample classified as benign is shown in equation 12.
\begin{align}
	D_{m_i} = dw_{m_i} - \bar{dw_{m_c}}
\end{align}
\begin{align}
	D_{b_i} = hw_{b_i} - \bar{dw_{b_c}}
\end{align}

The calibration set shown in Figure 1 is used to calculate the expected drift in the system; the expected drift is calculated to allow the system to know what changes in weighted distance values to expect. Expected drift $D_{e_m}$ for malign and $D_{e_b}$ for benign is calculated by taking the average drift in the calibration set and the average increase in drift $f$,  in datasets from different malign distributions. The average increase in drift $f$ is calculated by measuring the average drift in the training data and comparing the average drift in the samples showing increased drift in the calibration data. The calculation for the average drift for malign samples in the calibration set is shown in equation 14, and the average benign drift for benign predictions is shown in equation 15. The calculation for average drift in malign $\bar{D_m}$ and benign samples $\bar{D_b}$ in the training data is shown in equation 12 and 13.

\begin{align}
	\bar{D_m} = \frac{\sum_{i=1}^{n} D_{m_i}}{n}
\end{align}
\begin{align}
	\bar{D_b} = \frac{\sum_{i=1}^{n} D_{b_i} }{n}
\end{align}
Expected drift is calculated by combining the average drift in the calibration set and the average increase in drift for malign $f_m$ and benign samples $f_b$ from different malign and benign distributions, so in this case, the different distributions would be training and calibration sets; this is shown in equation 15 and equation 16 for expected drift in malign and benign samples, respectively. 
\begin{align}
	D_{e_m} = \bar{D_m} \cdot f_m
\end{align}
\begin{align}
	D_{e_b} = \bar{D_b} \cdot f_b
\end{align}
\subsubsection{Drift Boundary Thresholds}
The drift thresholds represent levels of drift that require different courses of action; the drift calibration layer allows us to define what is classified as normal and abnormal drift. Normal drift would require no action as it is within the framework's expected drift levels. Abnormal drift is beyond the standard threshold for drift but not beyond the maximum threshold. Abnormally drifting samples are recorded, and too many abnormal samples suggest a need to retrain. The drift calibration layer also defines drift boundaries; drift beyond the drift boundary would suggest a misclassification.
\begin{align}
	D_{f_{m} }= \frac{D_m}{D_{e_m}}
\end{align}
\begin{align}
	D_{f_{b}} = \frac{D_b}{D_{e_b}}
\end{align}
If the $D_{f_m}$ value for a sample classified as malign is above 1, this would mean the sample is showing abnormal drift for a malign classification; this is also the case for $D_{f_b}$ of a sample classified as benign. The abnormal drift threshold is decided by the expected drift value $D_e$. The expected drift value taking drift from many different distributions allows the abnormal drift threshold boundary to identify true outliers while not flagging samples that represent drift that can be expected. The threshold taking concept drift into account should extend the time between retraining while flagging samples displaying abnormal drift. Once $z$ number of samples have been identified as showing abnormal drift, the system needs to be retrained. The user defines the $z$ value.
\paragraph{\bf Maximum Drift}
The framework calculates the drift thresholds using the training and calibration sets. $D_{max}$ is a level of drift that a sample cannot exceed. Exceeding the $D_{max}$ would suggest an incorrect prediction, so a benign sample displaying drift beyond $D_{max}$  would be considered malign and vice versa for samples classified as benign. $D_{max}$  is defined as the point at which a sample, according to the weighted distance metric, is closer to the opposite class it is classified as. An example of this scenario would be a sample classified as benign by the underlying classifier, but the sample's drift level would exceed $D_{max}$, therefore it would be closer to malign than benign samples; it would be reclassified as malign. $D_{max}$ is calculated based on the training and incorrect classifications in the calibration set.$D_{max}$ is calculated using the training set shown in equation 19. The midpoint is defined and is $D_{max}$. If the drift of a sample is beyond $D_{max}$, then it will have a $D_{f_{max}}$ greater than 1; this is determined by equation 20.
\begin{align}
	D_{max}= \frac{\bar{hw_{r_c}} - \bar{hw_{b_c}}}{2}
\end{align}
\begin{align}
	D_{f_{max}} = \frac{D_i}{D_{max}}
\end{align}

\subsubsection{HEOM (Heterogeneous Euclidean Overlap Metric)}
The Heterogeneous Euclidean overlap metric (HEOM) is a distance metric designed to measure distances between two data samples and can work with both continuous and nominal attributes. The HEOM metric measures the distance between two values $x$ and $y$ given an attribute $a$. The operation of Heterogeneous Euclidean Overlap Metric is shown below in Equation 22 and 23. The HEOM metric is used for these experiments due to its general reliability and consistency in generating results in previous ransomware detection systems [25].

\begin{align}
	d_{heom}(x_1,x_2) = \sqrt{\sum_{a=1}^{m} heom^2_a(x_1,x_2)} 
	\label{equation 1}
\end{align}
Where
\begin{align}
	heom_a(x_1,x_2) = \frac{|x_{1a} - x_{2a}|}{range_a}
	\label{equtaion 2}
\end{align}

\subsection{Data}

We used datasets spanning 2016 to 2024; this approach was used to simulate concept-drift conditions in which the classifier is expected to detect samples it has not seen, which may have different or evolved characteristics from the training data. The datasets used are from large phishing dataset repositories [26], [27], [28], and [29]. Each dataset has a varying number of phishing and benign URLs; therefore, we have taken circa 10,000 samples from each dataset to ensure we obtain proportionate results and can test in proportion to the amount of training data we have.
\section{Experimental Setup}
The framework uses machine learning algorithms and similarity measures to aid the classification of malware samples under concept drift.
We configure machine learning algorithms and similarity measures to aid the classification of malware samples under concept drift. The initial setup for these experiments involved integrating the chosen datasets into the framework. The approach we use was designed to be dataset- and machine-learning-agnostic, supporting different distance and similarity metrics. We have chosen the Random Forest, Bayesian Network, and the Multi-Layer Perceptron, combined with the HEOM (Heterogeneous Euclidean Overlap Metric), for this batch of experiments due to their consistency and high performance in previous experiments [6]. We use standard machine-learning evaluation metrics to evaluate framework performance. The metrics we use to evaluate performance are shown in Table 1.

\begin{table}[h]
	\caption{Performance Metrics}
	\label{tab:title}
	\footnotesize
	\centering
	\begin{tabular}{|c|c|}
		\hline
		\textbf{Metric} & \textbf{Calculation} \\
		\hline
		
		TPR (True Positive Rate) / Recall 
		& $\frac{TP}{TP+FN}$ \\
		\hline
		
		False Positive Rate (FPR)
		& $\frac{FP}{FP+TN}$ \\
		\hline
		
		False Negative Rate (FNR)
		& $1-\text{TPR}$ \\
		\hline
		
		Precision
		& $\frac{TP}{TP+FP}$ \\
		\hline
		
	\end{tabular}
\end{table}

We have used four datasets with URLs collected from 2016 to 2024. The training data was generated using the feature selection component of the framework, and the underlying Random Forest algorithm was then trained on it. The training, calibration, and testing data are designed to represent different distributions, which was made easier by the dataset design, which used a training-on-older-data/testing-on-newer-data format. An example of our approach is using 2016 data to train an algorithm and testing it on newer data, such as 2021 data. Our initial hypothesis was that the detection rates in the training dataset would degrade when the algorithm was tested on the test data without any aid from the mathematical framework. The results of these initial tests are shown in Table 2 below.
Based on the algorithm's performance under concept drift, we test whether the framework improves the detection rate under such conditions. Ideally, we want to see samples that the underlying algorithm would have misclassified be correctly reclassified.

In addition to improving the detection algorithm's effectiveness at handling concept drift in phishing, we investigate the average difference between phishing and benign samples, as this can provide insight into how distinguishable the samples are based on the features produced by the framework. We aim to accomplish this by observing the average drift per class; this task is relatively simple because we measure drift for malign samples relative to their distance from other malign samples, and we measure a benign sample's drift relative to its distance from malign samples. Doing this allows us to assess the quality of the datasets and how similar or dissimilar the two classes of URLs are based on the provided features.
\subsection{Technical Configuration}
The mathematical framework runs using a combination of Java Code and the WEKA API. We currently use Java 19, and the WEKA API is from WEKA 3.8.6. The framework is tested with three algorithms. The Random Forest algorithm uses an unlimited depth and 100 individual trees in its current configuration. The Bayesian Network uses a simple estimator and the K2 Search algorithm. The Multilayer Perceptron uses a learning rate of 0.3 and momentum of 0.2. The implementation for all distance metrics is written in Java.
\subsection{Features}
We use features identified as being effective for Phishing detection from two prominent research papers on Phishing with Machine Learning detection [30] and [31]. The feature pool used identifies various characteristics of URLs. Additionally, we have added features to the feature pools given in these research papers. There are 75 features in total, designed to analyse components of the URL, such as the domain, the full URL, network characteristics (e.g., ports), and redirection behaviour. Additionally, the features analyse the composition of URLs, such as the characters and digits they contain. 

\section{Results and Discussion}
This section discusses our results and the conclusions we draw from them. We initially tested the Random Forest Algorithm on each dataset without concept drift to assess its baseline performance. Once we know the algorithm's baseline performance, we can measure how well or poorly the mathematical framework mitigates performance degradation under concept drift. The results of the Random Forest performance is shown below.

\subsection{First Phase Experiments}
In our first phase of experiments we see that the random forest yields some consistent results, followed by the Bayesian Network and the Multi-Layer Perceptron. We use the  feature selection algorithm to select features for each algorithm and have run the test with a 10 fold cross validation approach. All three algorithms follow a similar pattern where the detection rate is higher for the 2016 data compared to the 2019 and 2021 data. The Random Forest has a noticeably higher performance than the Bayesian Network and the Multi-Layer Perceptron as it is the only algorithm to achieve a detection rate above 80\%. Additionally, the Random Forest shows it is the most effective for false positives with the lowest average false positive rate across the three years of data. Next, we will observe how applying the mathematical framework can improve detection and false-positive rates when using the same three classification algorithms.

\begin{table}[H]
	\caption{Classifier Performance Comparison}
	\centering
	\scriptsize
	\setlength{\tabcolsep}{4pt}
	\begin{minipage}{0.31\textwidth}
		\centering
		\textbf{Random Forest} \\[4pt]
		\begin{tabular}{|c|p{0.58\linewidth}|}
			\hline
			\textbf{Dataset} & \centering\textbf{Performance Statistics}\arraybackslash \\
			\hline
			\textbf{2016} &
				\textbf{Detection:} 86.3\% \newline
				\textbf{FPR:} 10.8\% \newline
				\textbf{Precision:} 0.878 \newline
				\textbf{Recall:} 0.863 \\
			\hline
			\textbf{2019} &
				\textbf{Detection:} 83.1\% \newline
				\textbf{FPR:} 14.2\% \newline
				\textbf{Precision:} 0.845 \newline
				\textbf{Recall:} 0.831 \\
			\hline
			\textbf{2021} &
				\textbf{Detection:} 78.4\% \newline
				\textbf{FPR:} 16.9\% \newline
				\textbf{Precision:} 0.808 \newline
				\textbf{Recall:} 0.784 \\
			\hline
		\end{tabular}
	\end{minipage}
	\hfill
	\begin{minipage}{0.31\textwidth}
		\centering
		\textbf{Bayesian Network} \\[4pt]
		\begin{tabular}{|c|p{0.58\linewidth}|}
			\hline
			\textbf{Dataset} & \centering\textbf{Performance Statistics}\arraybackslash \\
			\hline
			\textbf{2016} &
				\textbf{Detection:} 79.3\% \newline
				\textbf{FPR:} 20.7\% \newline
				\textbf{Precision:} 0.786 \newline
				\textbf{Recall:} 0.793 \\
			\hline
			\textbf{2019} &
				\textbf{Detection:} 76.7\% \newline
				\textbf{FPR:} 28.3\% \newline
				\textbf{Precision:} 0.742 \newline
				\textbf{Recall:} 0.767 \\
			\hline
			\textbf{2021} &
				\textbf{Detection:} 78.4\% \newline
				\textbf{FPR:} 16.9\% \newline
				\textbf{Precision:} 0.808 \newline
				\textbf{Recall:} 0.784 \\
			\hline
		\end{tabular}
	\end{minipage}
	\hfill
	\begin{minipage}{0.31\textwidth}
		\centering
		\textbf{MLP} \\[4pt]
		\begin{tabular}{|c|p{0.58\linewidth}|}
			\hline
			\textbf{Dataset} & \centering\textbf{Performance Statistics}\arraybackslash \\
			\hline
			\textbf{2016} &
				\textbf{Detection:} 77.0\% \newline
				\textbf{FPR:} 18.0\% \newline
				\textbf{Precision:} 0.795 \newline
				\textbf{Recall:} 0.770 \\
			\hline
			\textbf{2019} &
				\textbf{Detection:} 74.2\% \newline
				\textbf{FPR:} 17.8\% \newline
				\textbf{Precision:} 0.784 \newline
				\textbf{Recall:} 0.742 \\
			\hline
			\textbf{2021} &
				\textbf{Detection:} 70.2\% \newline
				\textbf{FPR:} 18.0\% \newline
				\textbf{Precision:} 0.765 \newline
				\textbf{Recall:} 0.702 \\
			\hline
		\end{tabular}
	\end{minipage}
\end{table}

\subsection{Framework Experiments}

In our second-stage experiments, we observe how the mathematical framework affects the detection rate and false-positive rate under concept drift. We also observe how the number of samples considered showing "abnormal" drift changes when using drift values calculated by the drift calibration component of the framework.The three classification algorithms are combined with the framework's Heterogeneous Euclidean Overlap Metric. Tables V to VII cover the results of the framework experiments. The table structure shows the dataset year the algorithm was trained on, and the Performance Statistics column shows the algorithm's performance on the newer test data, as well as its performance when used with the mathematical framework. Table V presents the results we obtained using the Random Forest framework. We observe that the framework improves detection and reduces false positives across all scenarios, consistently lowering false-positive rates. The main observation is that the framework trained on 2016 data maintains a consistent detection rate across 2016, 2021 and 2024 data. The framework can maintain constant performance despite the degradation of the underlying algorithms' classification performance across multiple data distributions. We observe that the framework achieves higher performance with MLP and the Bayesian Network, although not as uniform or consistent as with the Random Forest. We observe that the framework shows higher and more erratic degradation when compared to the Bayesian Network and the Multi-Layer Perceptron in almost all experiments, but it provides a significant improvement in performance in most cases of heavy degradation. The framework does not uniformly help with the false positive rate; however, it cannot completely solve it, as the false positive rate remains relatively high. We hypothesise that the URL-based approach alone may have limitations due to features that create the potential for a higher false-positive rate, and the potential to degrade noticeably in terms of false-positive rate over time. In terms of abnormal reduction, we see the framework's calibration component reduces the number of samples marked as abnormal by circa 45\% across all experiments; this shows the framework is effectively able to compensate for roughly 45\% of the concept drift shown when the algorithms are exposed to samples from a different time than their training data. Naturally, we observe that samples are considered abnormal in the test data, as they have evolved beyond the training data to evade detection systems. The reduction in abnormal detections shows that calibration data helps expand the framework's understanding of what is evolving normally and what is truly abnormal to an unexpected degree.

\begin{table}[H]
	\caption{Experiments with Random Forest / MLP and Mathematical Framework}
	\setlength{\tabcolsep}{2pt}
	\centering
	\scriptsize
	\begin{minipage}{0.48\textwidth}
		\centering
		\textbf{Random Forest} \\[4pt]
		\begin{tabular}{|>{\centering\arraybackslash}m{0.22\linewidth}|p{0.72\linewidth}|}
			\hline
			\textbf{Training Year} & \textbf{Performance Statistics} \\
			\hline
			
			\textbf{2016} &
			
					\textbf{2019 Detection:} 56.2\% \newline
					\textbf{2019 FPR:} 21.4\% \newline
					\textbf{2019 Framework Detection:} 74.3\% \newline
					\textbf{2019 Framework FPR:} 9.1\% \newline
					\textbf{2021 Detection:} 65.9\% \newline
					\textbf{2021 FPR:} 33.4\% \newline
					\textbf{2021 Framework Detection:} 75.2\% \newline
					\textbf{2021 Framework FPR:} 19.8\% \newline
					\textbf{2024 Detection:} 56.9\% \newline
					\textbf{2024 FPR:} 19.3\% \newline
					\textbf{2024 Framework Detection:} 74.1\% \newline
					\textbf{2024 Framework FPR:} 10.3\% \newline
					\textbf{Avg Abnormal Reduction:} 47\%
			\\
			\hline
			
			\textbf{2019} &
			
					\textbf{2021 Detection:} 75.9\% \newline
					\textbf{2021 FPR:} 21.5\% \newline
					\textbf{2021 Framework Detection:} 84.0\% \newline
					\textbf{2021 Framework FPR:} 13.7\% \newline
					\textbf{2024 Detection:} 68.1\% \newline
					\textbf{2024 FPR:} 19.7\% \newline
					\textbf{2024 Framework Detection:} 78.6\% \newline
					\textbf{2024 Framework FPR:} 16.4\% \newline
					\textbf{Avg Abnormal Reduction:} 47\%
			\\
			\hline
			
			\textbf{2021} &
			
					\textbf{2024 Detection:} 68.9\% \newline
					\textbf{2024 FPR:} 19.7\% \newline
					\textbf{2024 Framework Detection:} 81.4\% \newline
					\textbf{2024 Framework FPR:} 18.2\% \newline
					\textbf{Avg Abnormal Reduction:} 42\%
			\\
			\hline
			
		\end{tabular}
	\end{minipage}
	\hfill
	\begin{minipage}{0.48\textwidth}
		\centering
		\textbf{MLP} \\[4pt]
		\begin{tabular}{|>{\centering\arraybackslash}m{0.22\linewidth}|p{0.72\linewidth}|}
			\hline
			\textbf{Training Year} & \textbf{Performance Statistics} \\
			\hline
			
			\textbf{2016} &
			
					\textbf{2019 Detection:} 48.2\% \newline
					\textbf{2019 FPR:} 9.8\% \newline
					\textbf{2019 Framework Detection:} 71.3\% \newline
					\textbf{2019 Framework FPR:} 9.8\% \newline
					\textbf{2021 Detection:} 51.2\% \newline
					\textbf{2021 FPR:} 18.4\% \newline
					\textbf{2021 Framework Detection:} 76.9\% \newline
					\textbf{2021 Framework FPR:} 9.8\% \newline
					\textbf{2024 Detection:} 55.3\% \newline
					\textbf{2024 FPR:} 21.8\% \newline
					\textbf{2024 Framework Detection:} 68.3\% \newline
					\textbf{2024 Framework FPR:} 13.1\% \newline
					\textbf{Avg Abnormal Reduction:} 47\%
			\\
			\hline
			
			\textbf{2019} &
			
					\textbf{2021 Detection:} 60.8\% \newline
					\textbf{2021 FPR:} 23.7\% \newline
					\textbf{2021 Framework Detection:} 83.6\% \newline
					\textbf{2021 Framework FPR:} 10.4\% \newline
					\textbf{2024 Detection:} 73.1\% \newline
					\textbf{2024 FPR:} 52.6\% \newline
					\textbf{2024 Framework Detection:} 85.1\% \newline
					\textbf{2024 Framework FPR:} 18.7\% \newline
					\textbf{Avg Abnormal Reduction:} 47\%
			\\
			\hline
			
			\textbf{2021} &
			
					\textbf{2024 Detection:} 74.7\% \newline
					\textbf{2024 FPR:} 28.4\% \newline
					\textbf{2024 Framework Detection:} 83.6\% \newline
					\textbf{2024 Framework FPR:} 13.1\% \newline
					\textbf{Avg Abnormal Reduction:} 42\%
			\\
			\hline
			
		\end{tabular}
	\end{minipage}
\end{table}

\begin{table}[H]
	\caption{Experiments with Bayesian Network and Mathematical Framework / Framework Experiments with Random Forest}
	\centering
	\scriptsize
	\setlength{\tabcolsep}{2pt}
	\begin{minipage}{0.48\textwidth}
		\centering
		\textbf{Bayesian Network} \\[4pt]
		\begin{tabular}{|>{\centering\arraybackslash}m{0.22\linewidth}|p{0.72\linewidth}|}
			\hline
			\textbf{Training Year} & \textbf{Performance Statistics} \\
			\hline
			\textbf{2016} &
					\textbf{2019 Detection:} 59.1\% \newline
					\textbf{2019 FPR:} 17.0\% \newline
					\textbf{2019 Framework Detection:} 81.3\% \newline
					\textbf{2019 Framework FPR:} 10.3\% \newline
					\textbf{2021 Detection:} 58.1\% \newline
					\textbf{2021 FPR:} 41.9\% \newline
					\textbf{2021 Framework Detection:} 70.6\% \newline
					\textbf{2021 Framework FPR:} 18.7\% \newline
					\textbf{2024 Detection:} 66.1\% \newline
					\textbf{2024 FPR:} 33.9\% \newline
					\textbf{2024 Framework Detection:} 74.2\% \newline
					\textbf{2024 Framework FPR:} 17.6\% \newline
					\textbf{Avg Abnormal Reduction:} 47\% \\
			\hline
			\textbf{2019} &
					\textbf{2021 Detection:} 45.4\% \newline
					\textbf{2021 FPR:} 34.6\% \newline
					\textbf{2021 Framework Detection:} 79.9\% \newline
					\textbf{2021 Framework FPR:} 14.2\% \newline
					\textbf{2024 Detection:} 56.9\% \newline
					\textbf{2024 FPR:} 32.8\% \newline
					\textbf{2024 Framework Detection:} 71.8\% \newline
					\textbf{2024 Framework FPR:} 19.9\% \newline
					\textbf{Avg Abnormal Reduction:} 47\% \\
			\hline
			\textbf{2021} &
					\textbf{2024 Detection:} 56.9\% \newline
					\textbf{2024 FPR:} 32.8\% \newline
					\textbf{2024 Framework Detection:} 71.8\% \newline
					\textbf{2024 Framework FPR:} 19.9\% \newline
					\textbf{Avg Abnormal Reduction:} 42\% \\
			\hline
		\end{tabular}
	\end{minipage}
	\hfill
	\begin{minipage}{0.48\textwidth}
		\centering
		\textbf{Random Forest (External Datasets)} \\[4pt]
		\begin{tabular}{|>{\centering\arraybackslash}m{0.22\linewidth}|p{0.72\linewidth}|}
			\hline
			\textbf{Dataset} & \textbf{Performance Statistics} \\
			\hline
			\textbf{Kaggle [32]} &
					\textbf{RF Detection:} 96.5\% \newline
					\textbf{Accuracy:} 96.7\% \newline
					\textbf{FPR:} 3.5\% \newline
					\textbf{Precision:} 0.964 \newline
					\textbf{Recall:} 0.965 \newline
					\textbf{Framework Detection:} 98.3\% \newline
					\textbf{Framework FPR:} 3.0\% \\
			\hline
			\textbf{Maribor [33]} &
					\textbf{RF Detection:} 93.9\% \newline
					\textbf{Accuracy:} 95.4\% \newline
					\textbf{FPR:} 6.1\% \newline
					\textbf{Precision:} 0.939 \newline
					\textbf{Recall:} 0.939 \newline
					\textbf{Framework Detection:} 97.8\% \newline
					\textbf{Framework FPR:} 3.1\% \\
			\hline
			\textbf{Badji Mokhtar [34]} &
					\textbf{RF Detection:} 94.4\% \newline
					\textbf{Accuracy:} 96.2\% \newline
					\textbf{FPR:} 3.0\% \newline
					\textbf{Precision:} 0.954 \newline
					\textbf{Recall:} 0.944 \newline
					\textbf{Framework Detection:} 96.8\% \newline
					\textbf{Framework FPR:} 2.5\% \\
			\hline
		\end{tabular}
	\end{minipage}
\end{table}

\begin{table}[H]
	\caption{Framework Experiments with MLP and Bayesian Network}
	\centering
	\scriptsize
	\setlength{\tabcolsep}{2pt}
	\begin{minipage}{0.48\textwidth}
		\centering
		\textbf{MLP} \\[4pt]
		\begin{tabular}{|>{\centering\arraybackslash}m{0.22\linewidth}|p{0.72\linewidth}|}
			\hline
			\textbf{Dataset} & \textbf{Performance Statistics} \\
			\hline
			\textbf{Kaggle [32]} &
					\textbf{MLP Detection:} 60.7\% \newline
					\textbf{Accuracy:} 75.4\% \newline
					\textbf{FPR:} 9.1\% \newline
					\textbf{Precision:} 0.869 \newline
					\textbf{Recall:} 0.607 \newline
					\textbf{Framework Detection:} 88.5\% \newline
					\textbf{Framework FPR:} 7.7\% \\
			\hline
			\textbf{Maribor [33]} &
					\textbf{MLP Detection:} 94.5\% \newline
					\textbf{Accuracy:} 90.0\% \newline
					\textbf{FPR:} 12.6\% \newline
					\textbf{Precision:} 0.945 \newline
					\textbf{Recall:} 0.882 \newline
					\textbf{Framework Detection:} 97.9\% \newline
					\textbf{Framework FPR:} 4.6\% \\
			\hline
			\textbf{Badji Mokhtar [34]} &
					\textbf{MLP Detection:} 92.3\% \newline
					\textbf{Accuracy:} 93.9\% \newline
					\textbf{FPR:} 6.8\% \newline
					\textbf{Precision:} 0.931 \newline
					\textbf{Recall:} 0.923 \newline
					\textbf{Framework Detection:} 95.8\% \newline
					\textbf{Framework FPR:} 3.4\% \\
			\hline
		\end{tabular}
	\end{minipage}
	\hfill
	\begin{minipage}{0.48\textwidth}
		\centering
		\textbf{Bayesian Network} \\[4pt]
		\begin{tabular}{|>{\centering\arraybackslash}m{0.22\linewidth}|p{0.72\linewidth}|}
			\hline
			\textbf{Dataset} & \textbf{Performance Statistics} \\
			\hline
			\textbf{Kaggle [32]} &
					\textbf{BN Detection:} 96.7\% \newline
					\textbf{Accuracy:} 96.1\% \newline
					\textbf{FPR:} 4.6\% \newline
					\textbf{Precision:} 0.954 \newline
					\textbf{Recall:} 0.967 \newline
					\textbf{Framework Detection:} 97.1\% \newline
					\textbf{Framework FPR:} 3.9\% \\
			\hline
			\textbf{Maribor [33]} &
					\textbf{BN Detection:} 94.5\% \newline
					\textbf{Accuracy:} 87.0\% \newline
					\textbf{FPR:} 19.0\% \newline
					\textbf{Precision:} 0.852 \newline
					\textbf{Recall:} 0.975 \newline
					\textbf{Framework Detection:} 94.5\% \newline
					\textbf{Framework FPR:} 9.0\% \\
			\hline
			\textbf{Badji Mokhtar [34]} &
					\textbf{BN Detection:} 91.5\% \newline
					\textbf{Accuracy:} 93.0\% \newline
					\textbf{FPR:} 5.0\% \newline
					\textbf{Precision:} 0.948 \newline
					\textbf{Recall:} 0.911 \newline
					\textbf{Framework Detection:} 96.4\% \newline
					\textbf{Framework FPR:} 3.2\% \\
			\hline
		\end{tabular}
	\end{minipage}
\end{table}
\clearpage
\subsection{Dataset Quality Evaluation}
This subsection discusses our results for the dataset quality evaluation. We have used the drift metrics produced by the mathematical framework to evaluate the statistical difference between phishing and benign samples. We do this to assess how well a dataset's features distinguish between phishing and benign samples. We have chosen three prominent, large datasets with predefined features to assess. We choose to use only distance metrics to measure dataset quality, providing a more objective, statistical view of the data rather than relying on the confidence of individual machine learning algorithms. The overall drift difference between benign and phishing samples is calculated using the training datasets used for the framework concept drift experiments.
\newline 
\newline 
For dataset evaluation, we use three external datasets and run experiments using the approach used in the datasets we created; the datasets are split into three categories: training, calibration, and test. The training data was generated using the feature selection component of the framework, and the underlying Random Forest, Bayesian Network and Multi-Layer Perceptron were then trained on it. The training, calibration, and testing data are designed to represent different distributions within each dataset, which was made easier by the dataset design, and the data in each dataset were collected over at least one year. An example would be using training data from early 2020, calibrating with data from mid 2020, and using test data from late 2020; this is the approach used across all three datasets, which span 1 to 3 years. Our initial hypothesis was that the training dataset's detection rates would degrade when the algorithm was exposed to test data outside its distribution. 
Across Tables VIII to X, the mathematical framework consistently increases the detection rate and reduces the false-positive rate, regardless of the algorithm or dataset. False-positive rates are kept notably low as well. Considering the sizes of the three datasets, the improvements, despite being small percentage-wise, are significant in real-number terms. 
\newline
\newline
Analysing the results in Table XI, the Maribor dataset shows the greatest distinguishable difference between phishing and benign samples. The average drift difference is more than twice that of the samples in the Kaggle dataset and close to twice that of the Badji Mokhtar dataset. The framework's performance is consistently strong across all three datasets; the greatest percentage improvement in detection with the Random Forest occurs in the Maribor dataset, jumping from 93.9\% to 97.8\%. The increase in performance correlates positively with the significant difference in drift between the average phishing and benign samples in the Maribor dataset. The increase in performance on the Kaggle, Maribor, and Badji Mokhtar datasets is identical; however, the Badji Mokhtar dataset shows a slightly larger difference between the drift in the average phishing samples and the average benign sample. This observation is consistent with the random forest, which shows that under drift, the Random Forest's prediction probability relies more on statistical differences between benign and malicious phishing samples. The correlation between statistical significance and the framework's ability to improve detection is weaker when using the Bayesian network. The lack of a clear statistical difference between the Bayesian network and the Multi-layer Perceptron could be attributed to the complexity of the two algorithms and their reliance on statistical differences between samples to a degree that Random Forest and potentially other tree-based algorithms might.
\begin{table}[H]
	\caption{Dataset Quality Evaluation}
	\label{tab:dataset_quality}
	\centering
	\footnotesize
	\setlength{\tabcolsep}{4pt}
	\hspace*{-1cm}
	\begin{tabular}{|>{\centering\arraybackslash}m{0.22\linewidth}|p{0.72\linewidth}|}
		\hline
		\textbf{Dataset} & \textbf{Drift Difference Between Benign and Phishing Samples} \\
		\hline
		
		\textbf{Kaggle [32]} &
		
					\textbf{Avg Benign Drift:} 0.48 \newline
					\textbf{Avg Phishing Drift:} -0.05 \newline
					\textbf{Drift Difference:} 0.49 \\
		\hline
		
		\textbf{Maribor [33]} &
		
					\textbf{Avg Benign Drift:} 0.81 \newline
					\textbf{Avg Phishing Drift:} -0.35 \newline
					\textbf{Drift Difference:} 1.17 \\
		\hline
		
		\textbf{Badji Mokhtar [34]} &
		
					\textbf{Avg Benign Drift:} 0.38 \newline
					\textbf{Avg Phishing Drift:} -0.25 \newline
					\textbf{Drift Difference:} 0.63 \\
		\hline
		
	\end{tabular}
\end{table}

\clearpage
\section{Conclusion}
To conclude, the research shows that using similarity indexes in combination with Machine Learning-based phishing detection systems can harden these detection systems against the effects of concept drift in the phishing space. Additionally, we have also demonstrated that these similarity indexes can be used to evaluate how distinguishable benign and phishing samples are in particular datasets.

\section{Acknowledgements}
This research was sponsored by the School of Science and Technology, City, St George's University of London.


\bigskip
\noindent\begin{minipage}[t]{0.13\textwidth}
	\vspace{0pt}
	\includegraphics[width=\linewidth]{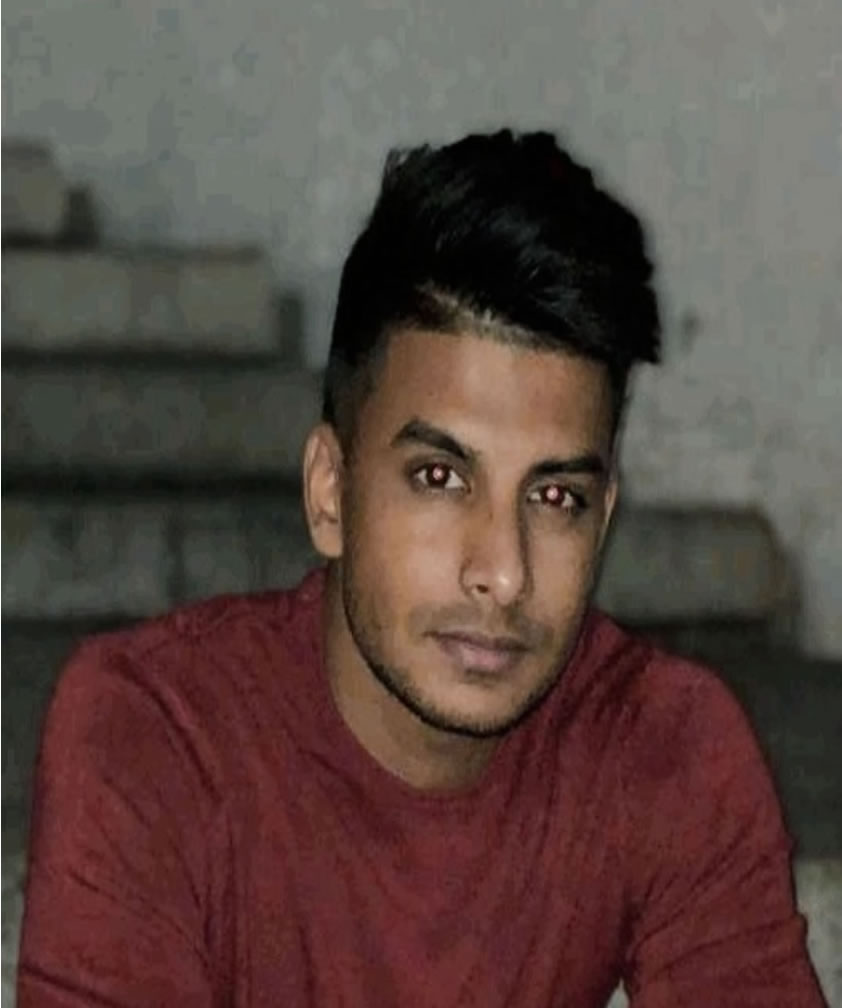}
\end{minipage}%
\hspace{0.02\textwidth}%
\begin{minipage}[t]{0.85\textwidth}
	\vspace{0pt}
	\noindent \textbf{Damien Warren Fernando} received a MSci in Computer Science and Cyber Security in 2017 from City, University of London. Having worked at City University of London as a Teaching assistant since late 2017, he is now a 4th year PhD student at City, University of London with an interest in researching Ransomware. While being a part time employee of City, University of London as a Teaching Assistant to the Cyber Security course, Damien provides teaching support along with managing and upgrading the Cyber Security penetration testing environment used by students.
\end{minipage}

\bigskip
\noindent\begin{minipage}[t]{0.13\textwidth}
	\vspace{0pt}
	\includegraphics[width=\linewidth]{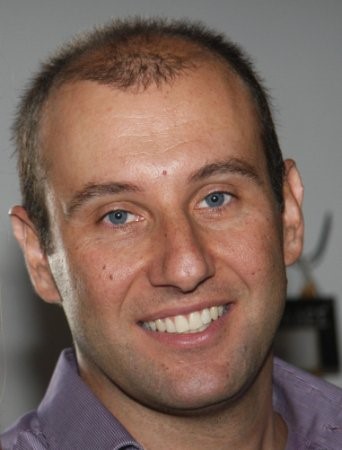}
\end{minipage}%
\hspace{0.02\textwidth}%
\begin{minipage}[t]{0.85\textwidth}
	\vspace{0pt}
	\noindent \textbf{Dr Nikos Komninos} received his PhD in 2003 from Lancaster University (UK) in Communication Systems with an Information Security focus. He is currently a Senior Lecturer (US System: Associate Professor) in Cyber Security in the Department of Computer Science at City University London. Part of his research has been patented and used in mobile phones by Telecommunication companies; in crypto-devices by Defense companies; and in healthcare applications by National Health Systems.
	Since 2000, he has participated, as a researcher or principal investigator, in many European and National R\&D projects in information security, systems and network security. He has authored and co-authored more than 80 journal publications, book chapters and conference proceedings publications in his areas of interest. He has been invited to give talks at conferences and Governmental Departments and train employees in Greek and UK businesses.
\end{minipage}

\end{document}